\documentclass[prc,showpacs,showkeys,superscriptaddress,floatfix]{revtex4}
\usepackage{graphicx,amssymb,amsmath}

\begin{document}
\title{On the relation between $E(5)-$models and the interacting boson
  model}

\author{Jos\'e Enrique Garc\'{\i}a-Ramos}
\email{enrique.ramos@dfaie.uhu.es} \affiliation{Departamento de
  F\'{\i}sica Aplicada, Universidad de Huelva, 21071 Huelva, Spain
}

\author{Jos\'e M. Arias} \email{ariasc@us.es}
\affiliation{Departamento de F\'{\i}sica At\'omica, Molecular y
  Nuclear, Facultad de F\'{\i}sica, Universidad de Sevilla,
  Apartado~1065, 41080 Sevilla, Spain
}

\begin{abstract}
  The connections between the $E(5)-$models (the original $E(5)$ using
  an infinite square well, $E(5)-\beta^4$, $E(5)-\beta^6$ and
  $E(5)-\beta^8$), based on particular solutions of the geometrical
  Bohr Hamiltonian with $\gamma$-unstable potentials, and the
  interacting boson model (IBM) are explored. For that purpose, the
  general IBM Hamiltonian for the $U(5)-O(6)$ transition line is used
  and a numerical fit to the different $E(5)-$models energies is
  performed, later on the obtained wavefunctions are used to calculate
  B(E2) transition rates. It is shown that within the IBM one can
  reproduce very well all these $E(5)-$models. The agreement is the
  best for $E(5)-\beta^4$ and reduces when passing through
  $E(5)-\beta^6$, $E(5)-\beta^8$ and $E(5)$, where the worst agreement
  is obtained (although still very good for a restricted set of lowest
  lying states).  The fitted IBM Hamiltonians correspond to energy
  surfaces close to those expected for the critical point.
  A phenomenon similar to the quasidynamical symmetry is observed.
\end{abstract}
\maketitle

\section{Introduction}
\label{sec-intro}
Both, the Bohr-Mottelson (BM) collective model
\cite{Bohr52,Bohr53,Bohr69} and the interacting boson model (IBM)
\cite{Arim76,Arim78,Scho78,Iach87} have thoroughly been used to study
the same kind of nuclear structure problems.  Although very different
in their formulation, both models present clear relationships.  In an
approximate way, the IBM can be interpreted as the second quantization
of the BM shape variables \cite{Jans74}.  More detailed connections
between both models were studied during the eighties by several
authors \cite{Diep80a,Diep80b,Gino80a,Gino80b,Kirs82,Gino82} and, more
recently, by Rowe and collaborators \cite{Rowe05}.  Both models have
three particular cases that can be easily solved and for which a clear
correspondence can be done. These three cases are: i) the BM
anharmonic vibrator and the dynamical symmetry $U(5)$ IBM limit, ii)
the BM $\gamma$-unstable deformed rotor and the dynamical $O(6)$ IBM
limit, and iii) the BM axial rotor and the dynamical symmetry $O(6)$
IBM limit including $Q\cdot Q\cdot Q$ interactions
\cite{Isac99,Rowe05}). Note that although it is traditionally accepted
the correspondence of the dynamical symmetry $SU(3)$ IBM limit to a
submodel of the BM, this fact has never been explicitly probed
\cite{Rowe05}.  Each of these cases are assigned to a particular shape
using the Hill-Wheeler variables $(\beta,\gamma)$ \cite{Hill53}:
spherical, deformed with $\gamma$-instability, and axially deformed,
respectively.  For transitional situations the correspondence between
the two models is difficult, as Rowe said ``what is simple in one
model will be complicated when expressed in terms of the observables
of the other''. This situation suggests, for the case of transitional
Hamiltonians, to look for the connection between BM and IBM through
numerical studies.

Among the transitional Hamiltonians, a specially interesting case
occurs when it describes a critical point in the transition from a
given shape to another. In general, for such a situation, where the
structure of the system can change abruptly by applying a small
perturbation, both, the BM and the IBM, have to be solved numerically.
However, recently Iachello has proposed schematic Bohr Hamiltonians
that intend to describe different critical points and that can be
solved exactly in terms of the zeros of Bessel functions. The first of
these models is known as $E(5)$ \cite{Iach00}.  $E(5)$ is designed to
describe the critical point at the transition from spherical to
deformed $\gamma$-unstable shapes. The potential to be used in the
differential Bohr equation is assumed to be $\gamma-$independent and,
for the $\beta$ degree of freedom an infinite square well is taken.
Similar models were proposed later on by Iachello,
called $X(5)$ and $Y(5)$ \cite{Iach01,Iach03}, to describe the
critical points between spherical and axially deformed shapes and
between axial and triaxial deformed shapes, respectively. All these
models give rise to spectra and electromagnetic transition rates that
are parameter free, up to a scale. In spite of their simplicity, some
experimental examples were found \cite{Cast00,Cast01}, just after the
appearance of these models.

In this work, we concentrate on $E(5)$ and related models. It will be
published elsewhere the corresponding study for $X(5)-$models
\cite{GarcXX}. The formulation of $E(5)$ attracted immediately
attention both experimentally and theoretically. Soon after the
introduction of the $E(5)$ model, the nucleus $^{134}$Ba was proposed
by Casten and Zamfir \cite{Cast00} as a realization of it. Other
experimental examples proposed are: $^{104}$Ru \cite{Fran01},
$^{102}$Pd \cite{Zamf02} and, $^{108}$Pd \cite{Zhan02}. Concerning
theoretical extensions of $E(5)$, first, Arias \cite{Aria01} proposed
a generalization of the E2 operator to be used with the $E(5)$ model,
then Caprio \cite{Capr02} checked that a substitution of the original
infinite well in the $\beta$ variable by a finite one, which makes the
model not exactly solvable anymore, provides similar results. It
showed that the $E(5)$ description is ``robust in nature'', {\it i.e.}
the main features of the model remain almost unchanged under strong
modification of the depth of the potential. Arias and collaborators
\cite{Aria03,Garc05} were the first authors who tried to analyze in a
quantitative way the connection between the $U(5)-O(6)$ IBM critical
point and the $E(5)$ model. In particular, they established, looking
to few observables that the IBM, at the critical point, gives results
close to $E(5)$ for a small ($N\approx 5$) number of bosons.  However,
the IBM results for large $N$ nicely reproduce the spectra and
electromagnetic transition rates of a Bohr Hamiltonian with a
$\beta^4$ potential (in the following $E(5)-\beta^4$). Once more, the
model is not analytically solvable anymore.  L\'evai and Arias
\cite{Leva04} solved the Bohr equation with a sextic potential with a
centrifugal barrier \cite{Ushv94}, arriving to almost closed
analytical formulae for the energies and wavefunctions.  Immediately
after, Bonatsos and collaborators explored the possibility of getting
numerical solutions for the $\gamma$-independent Bohr Hamiltonian with
potentials of the type $\beta^{2n}$, with $n\ge 1$ \cite{Bona04}.
These sequences of potentials allow to go from the vibrational limit,
$n=1$, to $E(5)$, $n \rightarrow \infty$. In particular, in
Ref.~\cite{Bona04} spectra and transition rates for the potentials
$\beta^4, \beta^6$ and, $\beta^8$, are given explicitly and compared
with the original $E(5)$ (infinite square well potential) case.  As
mentioned above, all these models are produced in the BM scheme and a
natural question is to ask for the corresponding equivalence in the
IBM. Is the IBM able for producing the same spectra and transition
rates? If yes, does the IBM Hamiltonian correspond to a critical
point? This work is intended to answer these questions for the $E(5)$
and related models ($-\beta^4, -\beta^6$ and, $-\beta^8$ potentials)
and analyze the convergence as a function of the boson number.

For that purpose, a large set of $E(5)$ and related models results for
excitation energies and transition rates are taken as reference for
numerical fits of the general $U(5)-O(6)$ IBM transitional Hamiltonian.
This procedure will allow to establish the IBM Hamiltonian which best
fit the different $E(5)-$models and their relation with the critical
points.

The paper is organized as follows: in section \ref{sec-fit} the
fitting procedure is described and the obtained results are commented.
Section \ref{sec-crit} is devoted to study the energy surfaces of the
fitted IBM Hamiltonians and to analyze these in relation to the
critical point. In section \ref{sec-quasi} the connection between the
present results and the concept of quasidynamical symmetry is
discussed. Finally, in section \ref{sec-conclu} the summary and
conclusions of this work are presented.

\section{The IBM fit to $E(5)-$models}
\label{sec-fit}
\subsection{The model}

The most general, including up to two-body terms, IBM Hamiltonian can
be written in multipolar form as,

\begin{eqnarray}
  \label{ham1}
  \hat H&=&\varepsilon_d \hat n_d +
  \kappa_0 \hat P^\dag \hat P
  +\kappa_1 \hat L\cdot \hat L+
  \kappa_2 \hat Q \cdot \hat Q+
  \kappa_3 \hat T_3\cdot\hat T_3 +\kappa_4 \hat T_4\cdot\hat T_4
\end{eqnarray}  
where $\hat n_d$ is the $d$ boson number operator, and
\begin{eqnarray}
  \label{P}
  \hat P^\dag&=&\frac{1}{2} ~ (d^\dag \cdot d^\dag - s^\dag \cdot s^\dag), \\
  \label{L}
  \hat L&=&\sqrt{10}(d^\dag\times\tilde{d})^{(1)},\\
  \label{Q}
  \hat Q&=& (s^{\dagger}\times\tilde d
  +d^\dagger\times\tilde s)^{(2)}-
  \frac{\sqrt{7}}{2}(d^\dagger\times\tilde d)^{(2)},\\
  \label{t3}
  \hat T_3&=&(d^\dag\times\tilde{d})^{(3)},\\
  \label{t4}
  \hat T_4&=&(d^\dag\times\tilde{d})^{(4)}.
\end{eqnarray} 
The symbol $\cdot$ stands for the scalar product, defined as $\hat
T_L\cdot \hat T_L=\sum_M (-1)^M \hat T_{LM}\hat T_{L-M}$ where $\hat
T_{LM}$ corresponds to the $M$ component of the operator $\hat T_{L}$.
The operator $\tilde\gamma_{\ell m}=(-1)^{m}\gamma_{\ell -m}$ (where
$\gamma$ refers to $s$ and $d$ bosons) is introduced to ensure the
correct tensorial character under spatial rotations.

The electromagnetic transitions can also be analyzed in the framework
of the IBM. In particular, in this work we will focus on the $E2$
transitions. The most general $E2$ transition operator including up to
one body terms can be written as,
\begin{equation}
  \label{te2}
  \hat T^{E2}_M=e_{eff}\left[(s^\dag \times 
    \tilde{d}+d^\dag\times\tilde{s})^{(2)}_M+
    \chi(d^\dag\times\tilde{d})^{(2)}_M\right],
\end{equation}
where $e_{eff}$ is the boson effective charge and $\chi$ is a
structure parameter.

The $E(5)-$models are intended to be of use for $\gamma$-unstable
nuclei having $O(5)$ as symmetry algebra.
For the construction of an IBM $\gamma$-unstable transitional
Hamiltonian it is sufficient to impose in Eq.~(\ref{ham1})
$\kappa_2=0$ (this implies that no Casimir operator from the $SU(3)$
algebra is included) as can be observed if the Hamiltonian
(\ref{ham1}) is rewritten in terms of Casimir operators (the
definition for the Casimir operators have been taken from
\cite{Fran94}):
\begin{eqnarray}
  \nonumber
  \hat H&=& \frac{\kappa_0}{4} N(N+4)+
  \big (\varepsilon_d +\frac{18}{35} \kappa_4 \big)\,\hat C_1[U(5)]+
  \frac{18}{35} \kappa_4 \,\hat C_2[U(5)]\\
  &+&\big(\kappa_1-\frac{\kappa_3}{10}-\frac{\kappa_4}{14}\big)\,\hat C_2[O(3)]+
  \big(\frac{\kappa_3}{2}-\frac{3}{14} \kappa_4\big)\,\hat C_2[O(5)]
  -\frac{\kappa_0}{4}\,\hat C_2[O(6)] .
  \label{ham-cas}
\end{eqnarray}
If additionally, we want to construct an IBM transitional Hamiltonian
that preserves the $O(5)$ symmetry, Casimir operators for $U(5)$,
$O(6)$ and $O(5)$ can be included but not the quadratic $O(3)$ Casimir
operator.  This condition, translated to the multipolar form language
used in Eq.~(\ref{ham1}), leads to the constraint
$\kappa_1-\kappa_3/10-\kappa_4/14=0$ (see Eq.~(\ref{ham-cas})).  In
addition, the structure parameter, $\chi$, in the $T^{E2}$ operator is
usually taken as zero in the standard IBM calculations for
$\gamma$-flat Hamiltonians.  In our calculations we will impose
$\kappa_2=0$, {\it i.e.}~the $\gamma$ flatness. To make more simple
the later analysis, we will restrict ourself to the case $\kappa_4=0$,
leaving as free parameters $\kappa_0$, $\kappa_1$, $\kappa_3$ (plus
$\varepsilon_d$ that fixes the energy scale). In practice, we do not
impose the constraint $\kappa_1-\kappa_3/10=0$ but, as it will be
shown, the condition will be fulfilled in every fit.

\subsection{The fitting procedure}
In this section we describe the procedure for getting the IBM
Hamiltonian which best fit the different $E(5)-$models.

The $\chi^2$ test is used to perform the fitting. The $\chi^2$
function is defined in the standard way,
\begin{equation}
  \label{chi2}
  \chi^2=\frac{1}{N_{data}-N_{par}}\sum_{i=1}^{N_{data}}\frac{(X_i
    (data)-X_i (IBM))^2}{\sigma_i^2},
\end{equation} 
where $N_{data}$ is the number of data, from a specific $E(5)$-model,
to be fitted,
$N_{par}$ is the number of parameters used in the IBM fit, $X_i(data)$
is an energy level (or a $B(E2)$ value) taken from a particular
$E(5)-$model, $X_i(IBM)$ is the corresponding calculated IBM value,
and $\sigma_i$ is an arbitrary error assigned to each $X_i(data)$.

In order to perform the fit, we minimize the $\chi^2$ function for the
energies, using
$\varepsilon_d$, $\kappa_0$, $\kappa_1$ and $\kappa_3$ as free
parameters and $\kappa_2$ and $\kappa_4$ fixed to zero.
For doing this task we use MINUIT \cite{minuit}, which allows to
minimize any multi-variable function.

The labels for the energy levels follow the usual notation introduced
for the $E(5)$ model: $\xi$ enumerates the zeros of the $\beta$ part
of the wave function, and $\tau$ is the label for the $O(5)$ algebra,
{\it i.e.}~the $O(5)$ seniority quantum number, which is a good
quantum number along all the transition from $U(5)$ to $O(6)$.  The
selected set of levels included in the fit for the different
$E(5)-$models are:
\begin{itemize}
\item For the $\xi=1$ band, all the states with angular momentum lower
  than $8$ and $\tau<5$. An arbitrary $\sigma= 0.001$ is used for
  these states except for the $2_1^+$ state for which $\sigma=0.0001$
  is used. This latter value allows to normalize all the IBM energies
  to $E(2_1^+)=1$. Note that the energy of the state $2_1^+$ is fixed
  arbitrarily to $1$ (remind that the spectrum is calculated up to a
  global scale factor).

\item For the $\xi=2$ band, all the states with angular momentum lower
  than $5$ and $\tau<3$. An arbitrary $\sigma=0.01$ is used for these
  states.
\item For the $\xi=3$ band, just the states with ($L=0,\tau=0$) and
  ($L=2,\tau=1$) are included. An arbitrary $\sigma=1$ is used for
  these states.
\end{itemize}
With this selection, the number of energy levels included in the fit,
$N_{data}$, is equal to $17$. Note that the state $0_1^+$ is not a
real data to be reproduced because we are interested just in
excitation energies and therefore the ground state is naturally fixed
to zero in both, $E(5)-$models and IBM. In Table \ref{tab-energ-fit}
the states included in the fit are explicitly given.

\begin{table}
  \begin{tabular}{|c|c|c|c|}
    \hline
    Band    &Error          & $\tau$   & States\\
    \hline
    $\xi=1$ &$\sigma=0.001$ & $\tau=0$ & $0_1^+$ \\
    &$\sigma=0.0001$& $\tau=1$ & $2_1^+$ \\
    &$\sigma=0.001$ & $\tau=2$ & $4_1^+, 2_2^+ $ \\
    &$\sigma=0.001$ & $\tau=3$ & $6_1^+, 4_2^+, 3_1^+, 0_3^+$ \\
    &$\sigma=0.001$ & $\tau=4$ & $6_2^+, 5_1^+, 4_3^+, 2_4^+$ \\
    \hline
    $\xi=2$ &$\sigma=0.01$ & $\tau=0$ & $0_2^+$ \\
    &$\sigma=0.01$ & $\tau=1$ & $2_3^+$ \\
    &$\sigma=0.01$ & $\tau=2$ & $4_4^+, 2_5^+ $ \\
    \hline
    $\xi=3$ &$\sigma=1$ & $\tau=0$ & $0_4^+$ \\
    &$\sigma=1$ & $\tau=1$ & $2_7^+$ \\
    \hline
  \end{tabular}
  \caption{States included in the energy fit.}
  \label{tab-energ-fit}
\end{table}

Once the IBM Hamiltonian is fixed for each $E(5)-$model by fitting the
energy levels, the $\chi^2$ function for the $B(E2)$ values is
constructed without any additional fitting. The only parameter in the
$E2$ operator (\ref{te2}), $e_{eff}$, is a global scale and is fixed
to give $B(E2; 2_1^+\rightarrow 0_1^+)=100$ in all cases (remind that
the structure parameter $\chi=0$ in the E2 operator for the
transitional class going from $U(5)$ to $O(6)$ studied here). The
transitions calculated are enlisted in Table \ref{tab-be2-fit}.

\begin{table}
  \begin{tabular}{|c|c|c|c|c||c|c|c|c|c|}
    \hline
    & $\xi_i$ &$\xi_f$ &$\tau_i$ &$\tau_f$& &$\xi_i$ &$\xi_f$ &$\tau_i$ &$\tau_f$\\ 
    \hline
    $B(E2:2_1^+\rightarrow  0_1^+)$& 1 & 1 & 1 & 0 &$B(E2:3_1^+\rightarrow  4_1^+)$& 1 & 1 & 3 & 2   \\                  
    $B(E2:4_1^+\rightarrow  2_1^+)$& 1 & 1 & 2 & 1 &$B(E2:0_3^+\rightarrow  2_2^+)$& 1 & 1 & 3 & 2   \\                  
    $B(E2:6_1^+\rightarrow  4_1^+)$& 1 & 1 & 3 & 2 &$B(E2:0_3^+\rightarrow  2_1^+)$& 1 & 1 & 3 & 1   \\                     
    $B(E2:2_2^+\rightarrow  2_1^+)$& 1 & 1 & 2 & 1 &$B(E2:2_3^+\rightarrow  0_2^+)$& 2 & 2 & 1 & 0   \\                   
    $B(E2:2_2^+\rightarrow  0_1^+)$& 1 & 1 & 2 & 0 &$B(E2:4_4^+\rightarrow  2_3^+)$& 2 & 2 & 2 & 1   \\                     
    $B(E2:4_2^+\rightarrow  2_1^+)$& 1 & 1 & 3 & 1 &$B(E2:2_7^+\rightarrow  0_4^+)$& 3 & 3 & 1 & 0   \\                    
    $B(E2:4_2^+\rightarrow  2_2^+)$& 1 & 1 & 3 & 2 &$B(E2:0_2^+\rightarrow  2_1^+)$& 2 & 1 & 0 & 1   \\                     
    $B(E2:4_2^+\rightarrow  4_1^+)$& 1 & 1 & 3 & 2 &$B(E2:0_4^+\rightarrow  2_3^+)$& 3 & 2 & 0 & 1   \\                    
    $B(E2:3_1^+\rightarrow  2_2^+)$& 1 & 1 & 3 & 2 & & & & &  \\                   
    \hline
  \end{tabular}
  \caption{$B(E2)$ transitions to be calculated.}
  \label{tab-be2-fit}
\end{table}

\subsection{The results}

We have done fits of the IBM Hamiltonian (\ref{ham1}) parameters, so
as to reproduce as well as possible the energies of the states given
in Table \ref{tab-energ-fit} and generated by the different
$E(5)-$models: $E(5)-\beta^4$, $E(5)-\beta^6$, $E(5)-\beta^8$, and
$E(5)$. Calculations for the four cited $E(5)-$models as a function of
the boson number, $N$, have been performed.

\begin{figure}[hbt]
  \centering
  \includegraphics[width=10cm]{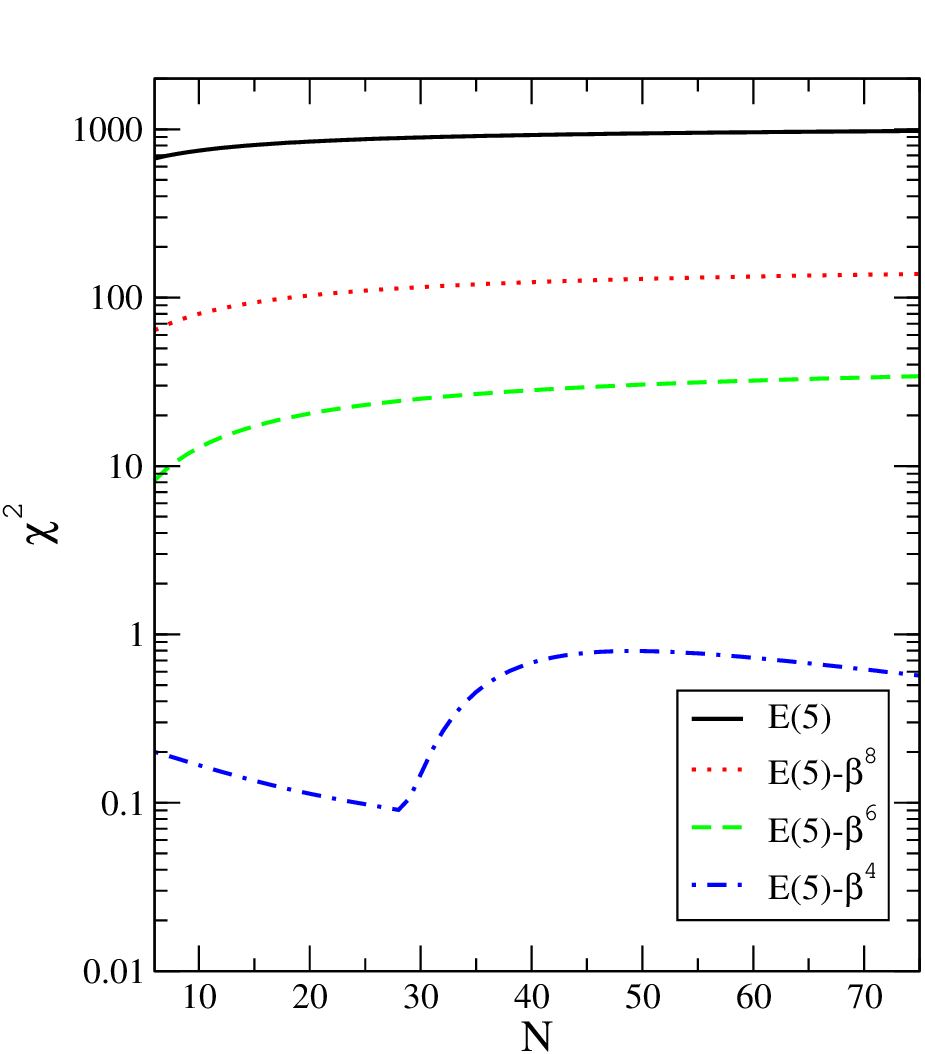}
  \caption{(color online). $\chi^2$ for the IBM fit to the energy
    levels of the different $E(5)$-models, as a function of $N$.}
  \label{fig-chi2-e5-not4t4}
\end{figure}

As it was mentioned before, $\kappa_2$ and $\kappa_4$ are set to zero
in Eq.  (\ref{ham1}), while $\varepsilon_d$, $\kappa_0$, $\kappa_1$
,and $\kappa_3$ are free parameters in a $\chi^2$ fit to the energy
levels produced by the different $E(5)-$models.  In figure
\ref{fig-chi2-e5-not4t4} the value of the $\chi^2$ for a best fit to
the different $E(5)-$models as a function of $N$ is shown. Different
type of lines in figure \ref{fig-chi2-e5-not4t4} represent the fit to
a different $E(5)-$model as stated in the legend box. It is clearly
observed that for any $N$ the agreement between the fitted IBM and the
$E(5)-\beta^4$ model is excellent and is getting worse for
$E(5)-\beta^6$, $E(5)-\beta^8$, up to reach $E(5)$ which is the worst
case. In particular $\chi^2(E(5)-\beta^4) \approx \chi^2(E(5))/1000$.
It is worth noting that these results change slowly with the boson
number and in all cases, except for $E(5)-\beta^4$ for which the
agreement is always excellent, the $\chi^2$ value is an increasing
function of $N$.

\begin{figure}[hbt]
  \centering
  \includegraphics[width=10cm]{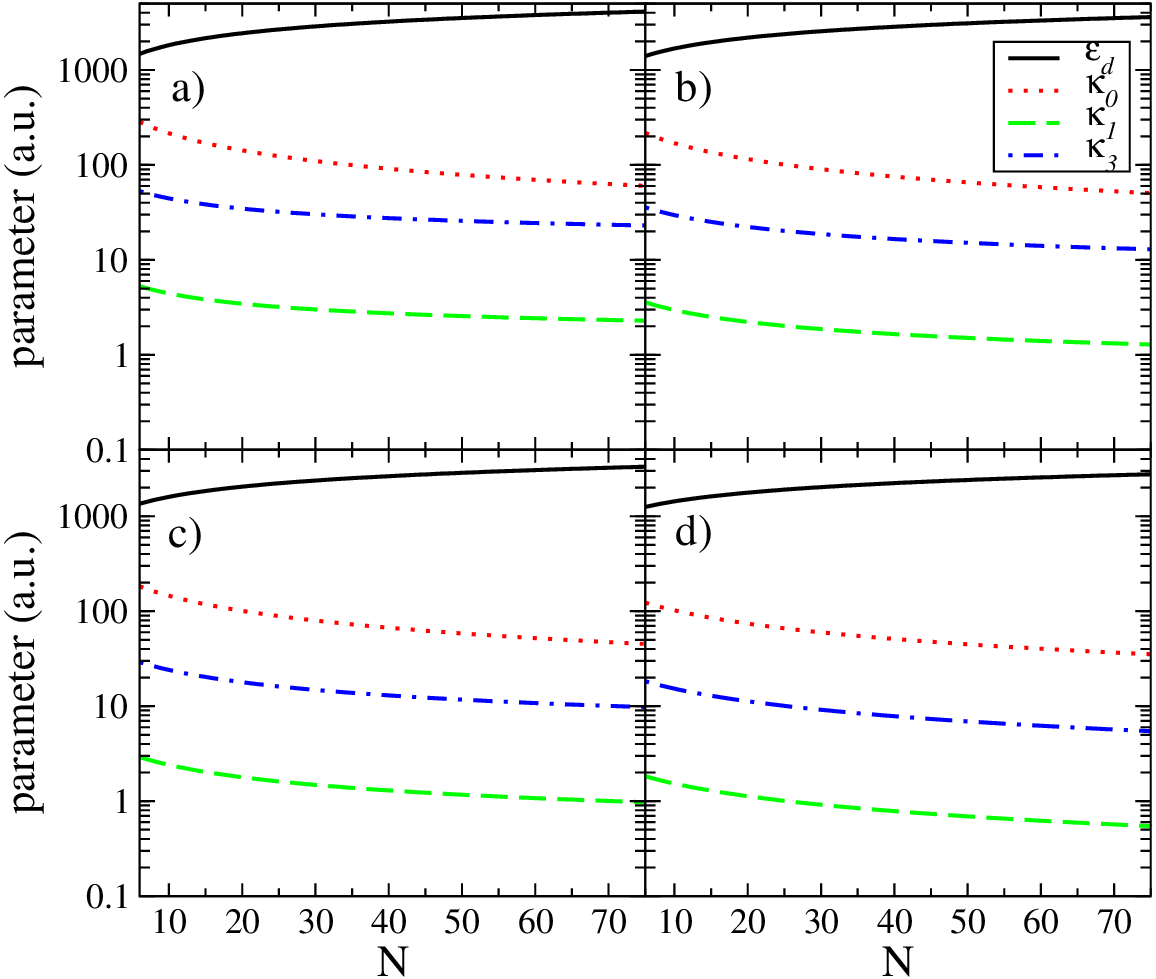}
  \caption{(color online). Values of the fitted IBM parameters (see
    text) as a function of $N$.  Different panels correspond to the
    fit to the different $E(5)-$models: a) $E(5)$, b) $E(5)-\beta^8$,
    c) $E(5)-\beta^6$, and d) $E(5)-\beta^4$.}
  \label{fig-par-not4t4}
\end{figure}

In figure \ref{fig-par-not4t4} the variation of the parameters fitted
in the Hamiltonian are shown. Note that the best fit parameters give
rise approximately to the cancellation of the quadratic Casimir
operator for $O(3)$, {\it i.e.}  $\kappa_1\approx \kappa_3/10$. This
can be quantitatively observed in Table \ref{table-parameters-not4t4}.

\begin{table}
  \begin{tabular}{|c|c|c|c|c|}
    \hline
    &$\varepsilon_d$ &$\kappa_0$ &$\kappa_1$&$\kappa_3$\\
    \hline
    $E(5)$        &3780.90&    69.74&   2.4308&  24.4520\\
    $E(5)-\beta^8$&3319.20&    58.26&   1.4028&  14.0770\\ 
    $E(5)-\beta^6$&3061.10&    52.06&   1.0753&  10.7760\\
    $E(5)-\beta^4$&2561.50&    40.24&   0.6218&   6.2157\\
    \hline
  \end{tabular}
  \caption{Parameters of the IBM Hamiltonians  
    used in table  \ref{table-states-not4t4}.} 
  \label{table-parameters-not4t4}
\end{table}

To have a clearer idea of the degree of agreement between the fitted
IBM results with the data from the $E(5)-$models, numerical
comparisons are shown in Table \ref{table-states-not4t4} for $N=60$.
This table includes not only the states used in the fit, but also an
extra set of states not included in it. These allow to control the
goodness of the obtained fit since they are predicted states which, as
we can see, have their counterpart in the $E(5)-$models.  The
agreement for $E(5)-\beta^4$, $E(5)-\beta^6$, and $E(5)-\beta^8$ is
really remarkable for all the states.  In the case of $E(5)$, only the
$\xi=1$ band is perfectly reproduced while for the bands with $\xi=2$
and $\xi=3$ the agreement is poor.

\begin{table}
  \begin{tabular}{|c|c||c|c||c|c||c|c||c|c|}
    \hline
    &$\xi,\tau$&E(5)&IBM&E(5)-$\beta^8$&IBM&E(5)-$\beta^6$&IBM&E(5)-$\beta^4$&IBM\\
    \hline 
    $0_1^+$     &1,0& 0.000& 0.000& 0.000&0.000&0.000&0.000&0.000&0.000\\
    $2_1^+$     &1,1& 1.000& 1.000& 1.000&1.000&1.000&1.000&1.000&1.000\\
    $4_1^+$     &1,2& 2.199& 2.214& 2.157&2.164&2.135&2.139&2.093&2.092\\
    $2_2^+$     &1,2& 2.199& 2.214& 2.157&2.164&2.135&2.139&2.093&2.092\\
    $0_2^+$     &2,0& 3.031& 3.051& 2.756&2.763&2.619&2.622&2.390&2.390\\
    $6_1^+$     &1,3& 3.590& 3.608& 3.459&3.467&3.391&3.395&3.265&3.265\\
    $4_2^+$     &1,3& 3.590& 3.608& 3.459&3.467&3.391&3.395&3.265&3.265\\
    $3_1^+$     &1,3& 3.590& 3.609& 3.459&3.467&3.391&3.395&3.265&3.265\\
    $0_3^+$     &1,3& 3.590& 3.609& 3.459&3.467&3.391&3.395&3.265&3.265\\
    $2_3^+$     &2,1& 4.800& 4.509& 4.255&4.148&4.012&3.961&3.625&3.632\\
    $6_2^+$     &1,4& 5.169& 5.159& 4.894&4.890&4.757&4.755&4.508&4.508\\
    $5_1^+$     &1,4& 5.169& 5.159& 4.894&4.890&4.757&4.755&4.508&4.508\\
    $4_3^+$     &1,4& 5.169& 5.159& 4.894&4.890&4.757&4.755&4.508&4.508\\
    $2_4^+$     &1,4& 5.169& 5.160& 4.894&4.890&4.757&4.755&4.508&4.508\\
    $4_4^+$     &2,2& 6.780& 6.108& 5.874&5.636&5.499&5.387&4.918&4.934\\
    $2_5^+$     &2,2& 6.780& 6.109& 5.874&5.636&5.499&5.387&4.918&4.934\\
    $0_4^+$     &3,0& 7.577& 6.682& 6.364&6.073&5.887&5.752&5.153&5.175\\
    $2_7^+$     &3,1&10.107& 8.511& 8.269&7.754&7.588&7.348&6.563&6.604\\
    \hline
    $6_3^+$*    &1,5& 6.930& 6.850& 6.456&6.421&6.225&6.207&5.813&5.817\\
    $5_2^+$*    &1,5& 6.930& 6.850& 6.456&6.421&6.225&6.207&5.813&5.817\\
    $4_5^+$*    &1,5& 6.930& 6.850& 6.456&6.421&6.225&6.207&5.813&5.817\\
    $2_6^+$*    &1,5& 6.930& 6.850& 6.456&6.421&6.225&6.207&5.813&5.817\\
    $6_{6,4}^+$*&2,3& 8.967& 8.669& 7.607&7.222&7.075&6.895&6.266&6.295\\
    $4_{7,6}^+$*&2,3& 8.967& 8.669& 7.607&7.222&7.075&6.895&6.266&6.295\\
    $3_{3,2}^+$*&2,3& 8.967& 8.669& 7.607&7.222&7.075&6.895&6.266&6.295\\
    $0_{6,5}^+$*&2,3& 8.967& 8.669& 7.607&7.222&7.075&6.895&6.266&6.295\\
    $4_9^+$*    &3,2&12.854&10.437&10.274&9.509&9.363&9.007&8.015&8.078\\
    $2_9^+$*    &3,2&12.854&10.437&10.274&9.509&9.363&9.007&8.015&8.078\\
    \hline
  \end{tabular}
  \caption{Comparison of energy levels for fitted IBM Hamiltonians, 
    with $N=60$, compared with those provided by the $E(5)$-models 
    (see text). The asterisk
    marks states not included in the fitting procedure. In the states
    labeled with two sub-indexes, the first one corresponds to $E(5)$, while
    the second to the rest of models.}
  \label{table-states-not4t4}
\end{table}

\begin{figure}[hbt]
  \centering
  \includegraphics[width=10cm]{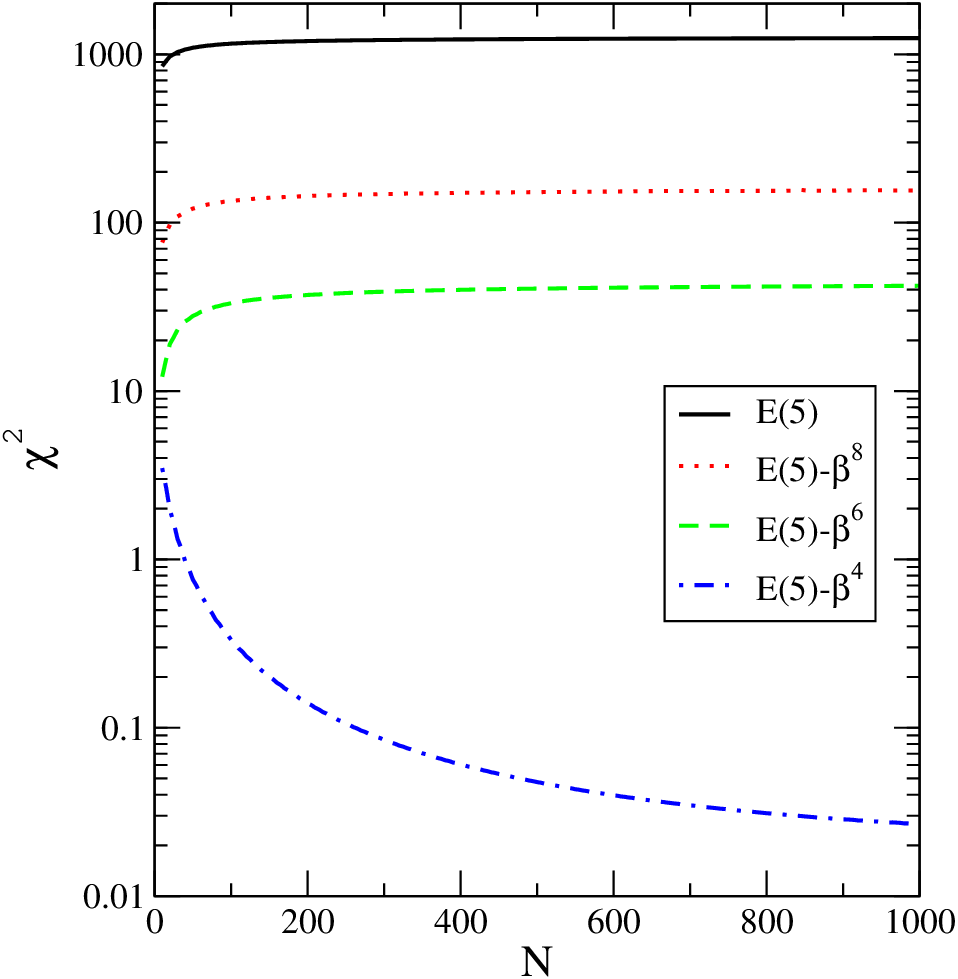}
  \caption{(color online). $\chi^2$ value for the IBM fit to the
    energy levels of the different $E(5)$-models, as a function of $N$
    (large $N$ limit), for an IBM Hamiltonian with $O(5)$ symmetry
    (see text).}
  \label{fig-chi2-e5-large}
\end{figure}

The IBM calculations presented are done with the usual IBM codes and
consequently are restricted, due to numerical limitations, to $N$
around 100.  However, one should note that for the transitional class
studied in this work
the $O(5)$ seniority is a good quantum number all along the
transition. This allows to diagonalize easily matrices corresponding
to large number of bosons using the procedure described in Ref.
\cite{Garc05} and explore the quality of the fits in the large N
limit.  The results of the $\chi^2$ fitting for such calculations are
presented in figure \ref{fig-chi2-e5-large} as a function of $N$. Note
that the curves presented in this figure do not match exactly with the
corresponding ones in Fig. \ref{fig-chi2-e5-not4t4} in the common $N$
range. This is because in the case in which the $O(5)$ symmetry is
imposed the $\chi^2$ function is constructed with only one state, of
those appearing in table \ref{tab-energ-fit}, per seniority. Then, the
number of states included in the fit is different, which results in 
slightly different values for the $\chi^2$ fitted function. The main
conclusion to be extracted from Fig.~\ref{fig-chi2-e5-large} is that
only the model $E(5)-\beta^4$ is exactly (at least for the states
considered in this work) reproduced by IBM Hamiltonians with $O(5)$
symmetry in the large N limit. For the rest of models the discrepancy
in the IBM fit slowly increases as a function of $N$.

\begin{figure}[hbt]
  \centering
  \includegraphics[width=10cm]{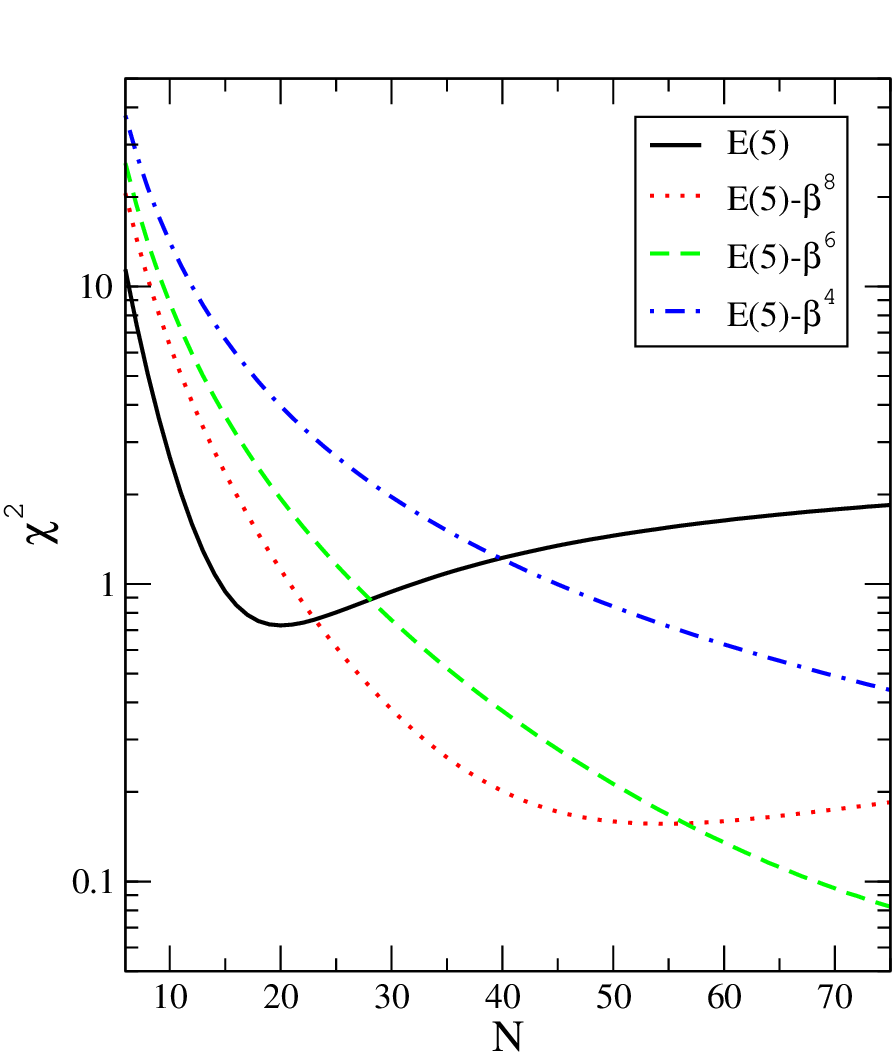}
  \caption{(color online). $\chi^2$ values for the $E2$ transition
    rates for the different $E(5)$-models, as a function of $N$, and
    an IBM electromagnetic operator $T(E2)=e_{\mbox{eff}} (s^\dag
    \tilde d+ d^\dag\tilde s)$.}
  \label{fig-be2-chi2-e5-not4t4}
\end{figure}

As a test for the produced wavefunctions with the fitted IBM
Hamiltonian, they are used for calculating $E2$ transition
probabilities, $B(E2)$. The effective charge (scale parameter) in the
$E2$ operator (\ref{te2}), is fixed so as to give
$B(E2;2_1^+\rightarrow 0_1^+)=100$, thus no free parameters are left
in this calculation. For the $B(E2)$'s calculated (not a fit) a
$\chi^2$ value has been obtained for each $E(5)-$model with an
arbitrary $\sigma=10$.  In figure \ref{fig-be2-chi2-e5-not4t4} the
corresponding $\chi^2$ value is plotted as a function of $N$ for all
the $E(5)-$models considered.  Figure \ref{fig-be2-chi2-e5-not4t4}
shows a clear dependence of $\chi^2$ on $N$. The $\chi^2$ value
decreases monotonically as $N$ increases for all the $E(5)-$models,
except for $E(5)$. In this last case, $\chi^2$ start increasing for
$N\approx 20$. For $N<20$, $E(5)$ provides the best agreement while
$E(5)-\beta^4$ is the worst. This fact changes when $N$ increases, and
for $N\approx 75$ already $E(5)-\beta^4$, $E(5)-\beta^6$, and
$E(5)-\beta^8$ provide a similar (excellent) agreement while the
$\chi^2$ value for $E(5)$ is clearly larger.

For a quantitative comparison, the $B(E2)$ values for the selected
transitions with $N=60$ are shown in table \ref{tab-be2-comp}. In this
table, it is clear the remarkable agreement between the IBM
calculations and $E(5)-$models. Note the $\Delta\tau=\pm 1$ selection
rule.  Thus, the wave functions produced by the fit to the energy
levels are giving roughly the correct $B(E2)$'s. However, it should be
noted that the calculated IBM $B(E2)$ values always increase as a
function of $N$. Therefore, looking at the transition rates
$B(E2:2_3^+\rightarrow 0_2^+)$, $B(E2:4_4^+\rightarrow 2_3^+)$,
$B(E2:2_7^+\rightarrow 0_4^+)$, $B(E2:0_2^+\rightarrow 2_1^+)$, and
$B(E2:0_4^+\rightarrow 2_3^+)$ in table \ref{tab-be2-comp} one
observes that, already for $N=60$, the IBM values are larger than
those provided by $E(5)$ and $E(5)-\beta^8$. This is also observed for
some transitions in the $E(5)-\beta^6$ model, but for none in the
$E(5)-\beta^4$ model. Thus, one expects for these models to start
giving larger $\chi^2$ values from a given $N$ value on. The IBM
results are always lower that the $E(5)-\beta^4$ ones and both are
approaching as $N$ increases.

In view of the excellent agreement between $E(5)-$models and the IBM,
we can state, that it is impossible to discriminate, from a
experimental point of view, between a $E(5)-$model and its IBM
counterpart.

\begin{table}
  \begin{tabular}{|r||r|r||r|r||r|r||r|r|}
    \hline
    & $E(5)$ & IBM & $E(5)-\beta^8$ &IBM  & $E(5)-\beta^6$ & IBM & $E(5)-\beta^4$ & IBM  \\
    \hline
    $B(E2:2_1^+\rightarrow  0_1^+)$&100   &100   &100    &100   &100    &100   &100    &100 \\
    $B(E2:4_1^+\rightarrow  2_1^+)$&167.4 &165.2 &173.3  &170.7 &176.6  &173.9 &183.2  &180.5 \\
    $B(E2:6_1^+\rightarrow  4_1^+)$&216.9 &215.4 &231.6  &227.0 &239.8  &233.9 &256.4  &248.7 \\
    $B(E2:2_2^+\rightarrow  2_1^+)$&167.4 &165.2 &173.3  &170.7 &176.6  &173.9 &183.2  &180.5 \\    
    $B(E2:2_2^+\rightarrow  0_1^+)$&0.0   &  0.0 &0.0    &  0.0 &0.0    &  0.0 &0.0    &  0.0 \\    
    $B(E2:4_2^+\rightarrow  2_1^+)$&0.0   &  0.0 &0.0    &  0.0 &0.0    &  0.0 &0.0    &  0.0 \\    
    $B(E2:4_2^+\rightarrow  2_2^+)$&113.6 &112.8 &121.3  &118.9 &125.6  &122.5 &134.3  &130.3 \\    
    $B(E2:4_2^+\rightarrow  4_1^+)$&103.3 &102.6 &110.3  &108.1 &114.2  &111.4 &122.1  &118.4 \\    
    $B(E2:3_1^+\rightarrow  2_2^+)$&154.9 &153.8 &165.5  &162.1 &171.3  &167.2 &183.1  &177.6 \\   
    $B(E2:3_1^+\rightarrow  4_1^+)$& 62.0 & 61.5 & 66.2  & 64.9 &68.5   & 66.8 &73.3   & 71.1 \\
    $B(E2:0_3^+\rightarrow  2_2^+)$& 216.9&215.4 & 231.6 &227.0 &239.8  &233.9 &256.4  &248.7 \\
    $B(E2:0_3^+\rightarrow  2_1^+)$&0.0   &  0.0 &0.0    &  0.0 &0.0    &  0.0 &0.0    &  0.0 \\
    $B(E2:2_3^+\rightarrow  0_2^+)$&75.2  & 90.2 &91.2   & 95.9 &99.0   & 99.6 &112.6  &107.9 \\ 
    $B(E2:4_4^+\rightarrow  2_3^+)$&124.3 &152.3 &156.1  &163.5 &172.0  &170.5 &197.9  &186.5 \\ 
    $B(E2:2_7^+\rightarrow  0_4^+)$&65.7  & 89.3 &91.6   & 97.9 &103.7  &103.4 &126.6  &115.9 \\ 
    $B(E2:0_2^+\rightarrow  2_1^+)$&86.8  & 81.6 &107.6  &100.8 &119.0  &112.1 &141.8  &135.4 \\ 
    $B(E2:0_4^+\rightarrow  2_3^+)$&123.2 &155.0 &178.5  &182.0 &205.3  &198.5 &257.9  &235.6 \\
    \hline
  \end{tabular}
  \caption{$B(E2)$ values obtained, for $N=60$, for fitted IBM
    Hamiltonians (see text) compared with those provided by the
    different $E(5)-$models.}
  \label{tab-be2-comp}
\end{table}

\section{The critical Hamiltonian}
\label{sec-crit}
One of the most attractive features of the $E(5)-$models treated in
this work is that they are supposed to describe, at different
approximation levels, the critical point in the transition from
spherical to deformed $\gamma$-unstable shapes. Since they are connected to a
given IBM Hamiltonian, as shown in the preceding section, this should
correspond to the critical point in the transition from $U(5)$ to
$O(6)$ IBM limits, {\it i.e.} this Hamiltonian should produce an
energy surface with ${\left(\frac{d^2
      E}{d\beta^2}\right)_{\beta=0}}=0$. Is this the case for the
fitted IBM Hamiltonians obtained in the preceding section?  Before
starting with the discussion it is necessary to establish a measure on
how close is a given IBM Hamiltonian to the critical point.

An energy surface can be associated to a given IBM Hamiltonian by
using the intrinsic state formalism \cite{Gino80b,Diep80a,Diep80b}
which introduces the shape variables $(\beta,\gamma)$ in the IBM. To
define the intrinsic state one has to consider that the dynamical
behavior of the system can be approximately described in terms of
independent bosons moving in an average field \cite{Duke84}. The
ground state of the system is written as a condensate, $|c\rangle$, of
bosons that occupy the lowest-energy phonon state, $\Gamma_c^\dag$:
\begin{equation}
  \label{GS}
  | c \rangle = \frac{1}{\sqrt{N!}} (\Gamma^\dagger_c)^N | 0 \rangle,
\end{equation}
where
\begin{equation}
  \label{bc}
  \Gamma^\dagger_c = \frac{1}{\sqrt{1+\beta^2}}~ \left (s^\dagger + \beta
    \cos     \gamma          \,d^\dagger_0          + \frac{1}{\sqrt{2}}~\beta
    \sin\gamma\,(d^\dagger_2+d^\dagger_{-2}) \right) .
\end{equation}
$\beta$ and $\gamma$ are variational parameters related with the shape
variables in the geometrical collective model \cite{Gino80b}.  The
expectation value of the Hamiltonian (\ref{ham1}) in the intrinsic
state (\ref{GS}) provides the energy surface of the system,
$E(N,\beta,\gamma)=\langle c|\hat H| c \rangle$.  This energy surface
in terms of the parameters of the Hamiltonian (\ref{ham1}) and the
shape variables can be readily obtained \cite{Isac81},
\begin{eqnarray}
  \label{Ener1}
  \langle c|\hat H | c \rangle&=&
  {\displaystyle {\frac{N\beta^2}{(1+\beta^2 )}}}
  \Bigl(\varepsilon_d+
  6 \,\kappa_1 
  -\frac{9}{4}\,\kappa_2+\frac{7}{5}\,\kappa_3+\frac{9}{5}
  \,\kappa_4\Bigr)\nonumber\\
  &+&{\displaystyle {\frac{N(N-1)}{{{(1+\beta^2)}^2}}}}
  \Big[\frac{\kappa_0}{4}+\beta^2(- \frac{\kappa_0}{2}+4\,\kappa_2)
  +2\,{\sqrt{2}}\,\beta^3\,\kappa_2\,\cos(3\,\gamma)
  \nonumber\\
  \qquad\qquad\qquad
  &&+\beta^4(\frac{\kappa_0}{4}
  +\frac{\kappa_2}{2}+\frac{18}{35}\,\kappa_4)\Big].
\end{eqnarray}

The shape of the nucleus is defined through the equilibrium value of
the deformation parameters, $\beta$ and $\gamma$, which are obtained
minimizing the ground state energy, $\langle c|\hat{H}|c \rangle$. A
spherical nucleus has a minimum in the energy surface at $\beta=0$,
while a deformed one presents the minimum at a finite value of
$\beta$. The parameter $\gamma$ represents the departure from axial
symmetry, {\it i.e.} $\gamma=0$ and $\gamma=\pi/6$ stand for an
axially deformed nucleus, prolate and oblate respectively, while any
other value corresponds to a triaxial shape. An additional situation
appears when the energy surface is independent on $\gamma$ but
presents a minimum in $\beta$, being the nucleus $\gamma$-unstable. It
should be noted that for a general IBM Hamiltonian including up to two
body terms the shape is either axially symmetric or $\gamma$-unstable.
Moreover, the Hamiltonians considered in this work correspond always
to the $\gamma$-unstable situation.

With the tools described above one can study phase transitions in the
IBM \cite{Diep80a}. First, the parameters that define the Hamiltonian
are the control parameters and normally are chosen in such a way that
only one of them is a variable, while the rest remain constant. The
deformation parameters $\beta$ and $\gamma$ become the order
parameters, although in our case the only order parameter is $\beta$.
Roughly speaking, a phase transition appears when there exists an
abrupt change in the shape of the system when changing smoothly the
control parameter. The phase transitions can be classified according
to the Ehrenfest classification \cite{Stan71}.  First order phase
transitions appear when there exists a discontinuity in the first
derivative of the energy with respect to the control parameter. This
discontinuity appears when two degenerate minima exist in the energy
surface for two values of the order parameter $\beta$. Second order
phase transitions appear when the second derivative of the energy with
respect to the control parameter displays a discontinuity. This
happens when the energy surface presents a single minimum for
$\beta=0$ and the surface satisfies the condition ${\left(\frac{d^2
      E}{d\beta^2}\right)_{\beta=0}}=0$. In a more modern
classification, second order phase transitions belongs to the high
order or continuous phase transitions \cite{Stan71}.

To determine whether a given Hamiltonian corresponds to a critical
point or not, the flatness or the existence of two degenerate minima
in the energy surface should be investigated.  For the case of one
parameter IBM Hamiltonian, {\it e.g.}~Consistent Q (CQF) Hamiltonians
\cite{Warn82}, it is simple to find an analytical expression for the
critical control parameter in the Hamiltonian. However, for a general
IBM Hamiltonian it is necessary to rewrite the energy surface in a
special way, as the one presented in Ref.~\cite{Lope96}. There, the
authors manage to write the energy surface of a general IBM
Hamiltonian in terms of two parameters. The authors make use of some
concepts from the Catastrophe Theory \cite{Gilm81} to define the two
essential parameters, ($r_1,r_2$). In terms of these they find
expressions for the locus, in the essential parameter space, that
gives a critical point at the origin in $\beta$, called bifurcation
set, and for the locus that gives rise to two degenerate minima,
called Maxwell set. For the Hamiltonians considered in section
\ref{sec-fit} $\kappa_2=0$ and $\kappa_4=0$, in these cases $r_2=0$
and $r_1$ can be written as,
\begin{equation}
  r_1=\frac{a_3-u_0+\tilde\varepsilon/(N-1)}{2 a_1+
    \tilde\varepsilon/(N-1)-a_3}, 
  \label{r1}
\end{equation}
where
\begin{eqnarray}
  \tilde\varepsilon&=& \varepsilon_d+
  6\,\kappa_1 +\frac{7}{5}\,\kappa_3 , \nonumber\\
  a_1&=&\frac{1}{4}\,\kappa_0,
  \nonumber\\
  a_3&=&-\frac{1}{2}\,\kappa_0 ,
  \nonumber\\
  u_0&=&\frac{\kappa_0}{2}.
  \label{coeff}
\end{eqnarray}

Note that in the large $N$ limit, $\varepsilon_d$ is proportional to
$N$ (see figure \ref{fig-par-not4t4}) and therefore (\ref{r1}) can be
approached by,
\begin{equation}
  r_1\approx\frac{\varepsilon_d/N-\kappa_0}{ \varepsilon_d/N+\kappa_0}~.
  \label{r1p}
\end{equation}
This expression agrees with the use of an energy surface derived
through a Holstein-Primakoff expansion \cite{Aria07}.
   
\begin{figure}[hbt]
  \centering
  \includegraphics[width=10cm]{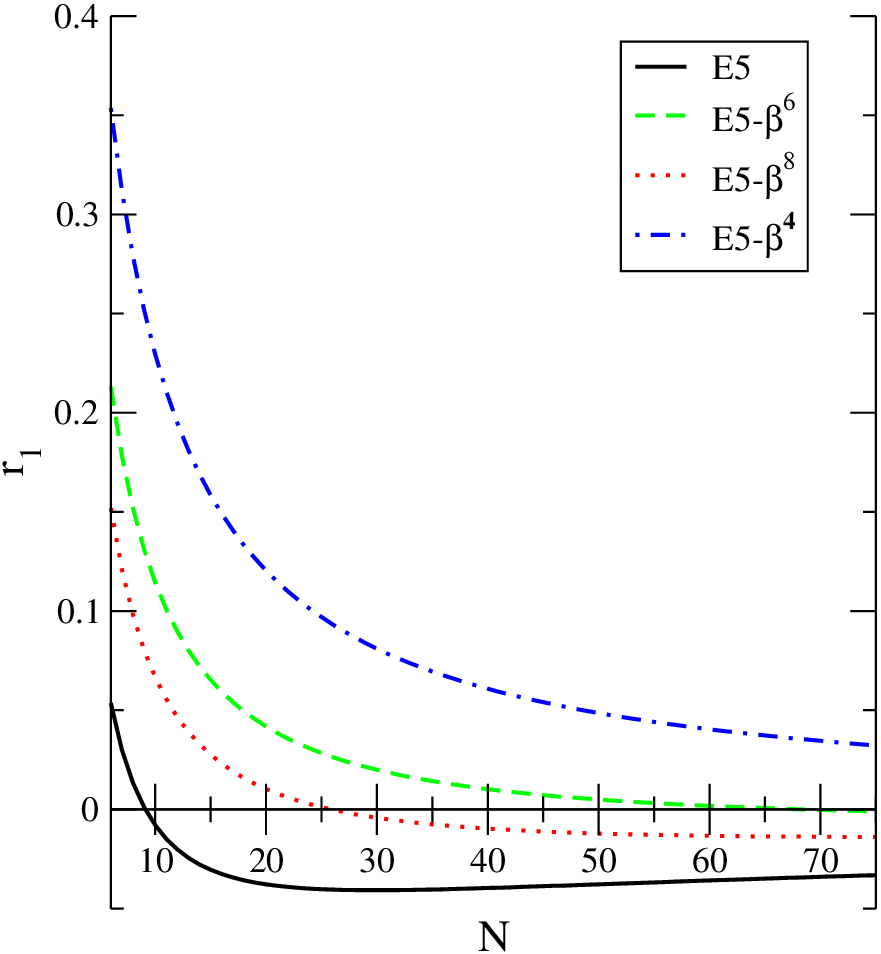}
  \caption{(color online). Values of $r_1$ (see text for definition)
    as a function of $N$ for the fitted IBM Hamiltonians.}
  \label{fig-r1-noT4T4}
\end{figure}

In this language, a critical Hamiltonian corresponds to $r_1=0$. In
figure \ref{fig-r1-noT4T4} the values of $r_1$ as a function of $N$
for the IBM Hamiltonians obtained from the fit are presented for the
different $E(5)-$models studied.  For the $E(5)$ model, the fitted IBM
Hamiltonian produces $r_1=0$ for $N\approx 7$.  In the case of the
$E(5)-\beta^8$ the value $r_1=0$ is obtained for $N \approx 25$, while
for $E(5)-\beta^6$ it is obtained for $N \approx 70$. For the
$E(5)-\beta^4$ model it is known that $r_1=0$ is reached for very
large number of bosons \cite{Aria03,Garc05}.

\begin{figure}[hbt]
  \centering
  \includegraphics[width=10cm]{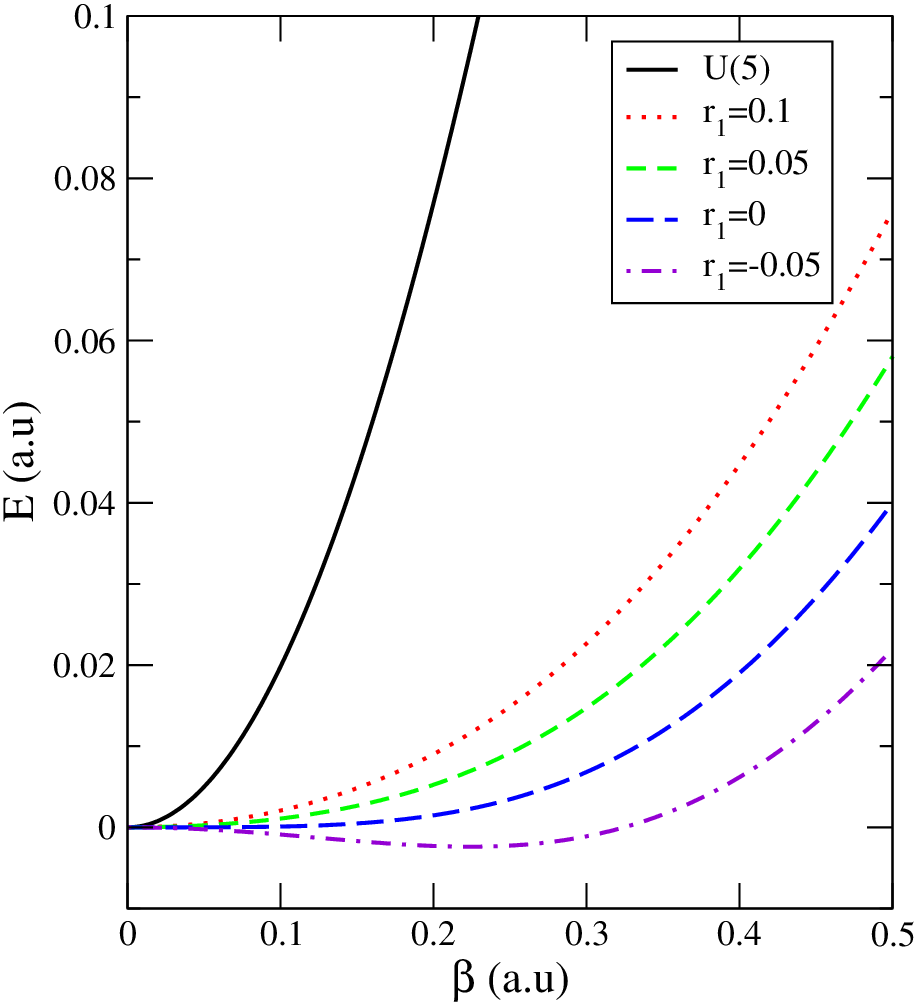}
  \caption{(color online). IBM energy surfaces as a function of
    $\beta$, for selected values of $r_1$ (see text for definition).}
  \label{fig-ener-r1}
\end{figure}

It is worth to show (see figure \ref{fig-ener-r1}) that for all the
fitted IBM Hamiltonians, the resulting energy surfaces are quite flat
in a large interval of $N$ values and, therefore, it is justified to
say that the fitted IBM Hamiltonians are very close to the critical
area. As a consequence, the $E(5)-$models will be appropriated to
describe phase transition regions close to the critical point.

\section{ Quasidynamical symmetries}
\label{sec-quasi}
The concept of quasidynamical symmetry (QDS) was introduced in
Refs.~\cite{Rowe04a,Rowe04b,Rowe04c,Rowe05,Rowe05b} and has been used
in the study of phase transitions.  This concept is very useful for
working with Hamiltonians that present as limits two dynamical
symmetries (depending on the value of a control parameter). In this
situation, the system shows the tendency to hold onto a given symmetry
until the control parameter reach a critical value, passing the
system, at this moment, onto the other symmetry. The remarkable
feature is that the system can present a set of states that behave as
belonging to irreducible representations (irreps) of the corresponding
symmetry group, although in fact, they do not belong to a given irrep
but to a mixture of them.

In mathematical terms, a QDS can be defined through the {\it embedded
  representations} \cite{Rowe04c}: ``If a subset of states of a system
are in one-to-one correspondence with the states of an irrep of a
group G and if all the properties of the subset of states associated
with observables in the Lie algebra of G (including their
relationships to one another but not necessarily their relationships
to states outside of the subset) are as they would be if the states
actually belonged to an irrep of the group G, then the subset of
states is said to span an embedded representation of G''. Therefore,
in the case of a QDS, there exist a set of states that behave as
belonging to a unique irrep of G, although that is only apparent,
because they correspond to a superposition of irreps, but all their
observables (up to certain degree of accuracy) are identical to the
ones of states within a given irrep. In summary, the states can be
expressed as a coherent superposition of irreps that behave as a
single one.  Note that to show that a QDS exists, one has to fix a
subset of states and the degree of accuracy for the comparison with
the observables of the dynamical symmetry.

In our comparison between the IBM and the $E(5)-$models we observe a
phenomenon which resembles the QDS, {\it i.e.}~part of the IBM
spectrum behaves as having $E(5)-$symmetry, although, indeed they do
not have such a symmetry. We should emphasize that this is not a real
QDS for two reasons: i) $E(5)-$cases are not dynamical symmetry limits
of the IBM and ii) the BM and the IBM have different Hilbert spaces.
Indeed, it is not possible to define irreps in $E(5)-$models and
therefore embedded representations. We will call this situation
quasi-critical point symmetry (QCPS) \cite{Rowe07}.

In order to study in detail the QCPS one has to fix the degree of
accuracy to be demanded to the observables.  In our study, for the
energies an accuracy of $1\%$ for all the states belonging to a given
$\xi$ is set while for the $B(E2)$ values an accuracy of $10\%$ for
all the studied intra-band transitions in a given $\xi$ is selected.

Tables \ref{table-states-not4t4} and \ref{tab-be2-comp}, which
correspond to $N=60$ are analyzed below,
\begin{itemize}
\item $E(5)$: only the states in the $\xi=1$ band present $E(5)$ QCPS.
\item $E(5)-\beta^8$: only the $\xi=1$ states present $E(5)-\beta^8$
  QCPS.
\item $E(5)-\beta^6$: only the $\xi=1$ states present $E(5)-\beta^6$
  QCPS.
\item $E(5)-\beta^4$: all the studied states, $\xi=1$, $\xi=2$ and
  $\xi=3$, present $E(5)-\beta^4$ QCPS.
\end{itemize}

These results, regarding the energies, can be extended to larger
values of $N$ too (see figure \ref{fig-chi2-e5-not4t4}), {\it i.e.}
the values of the energies remain stable when $N$ increases, while for
the $B(E2)$ values the observed differences become larger, specially
in the $E(5)$ case.

\section{Summary and conclusions}
\label{sec-conclu}
In this paper, we have studied the connection between the
$E(5)-$models and the IBM on the basis of a numerical mapping between
models. To establish the mapping we have performed a best fit of the
general $U(5)-O(6)$ transitional IBM Hamiltonian to a selected set of
energy levels produced by several $E(5)-$models.  Later on, a check to
the wavefunctions, obtained with the best fit parameters, has been
done by calculating relevant $B(E2)$ transition rates.  All
calculations have been done as a function of the number of bosons.
Once the best fit IBM Hamiltonians to the different $E(5)-$models are
obtained, their energy surfaces are constructed and analyzed with the
help of the Catastrophe Theory so as to know how close they are to a
critical point. Finally, the concept of quasi-critical point symmetry
is introduced, as similar to the idea of quasidynamical symmetry.

We have shown that it is possible, in all cases, to establish a
one-to-one mapping between the $E(5)-$models and the IBM with a
remarkable agreement for both the energies and the $B(E2)$ transition
rates.  In general, the goodness of the fit to the energies is
independent on the number of bosons, but the corresponding $B(E2)$
transition rates are indeed sensitive to $N$. This is so specially in
the $E(5)$, for which the $\chi^2$ value reaches a minimum for $N$
small ($N\approx 7$) and from there on increases notably as a function
of $N$. Globally, the best agreement is obtained for the
$E(5)-\beta^4$ Hamiltonian and the worst for the $E(5)$ case.
For the case of very large number of bosons and Hamiltonians with
$O(5)$ symmetry we have confirmed the results of \cite{Aria03,Garc05},
{\it i.e.} the only $E(5)-$model that can be reproduced exactly by the
IBM is $E(5)-\beta^4$, corresponding such a Hamiltonian with the
critical point of the model ($r_1=0$). A consequence of this excellent
agreement is that it is impossible, from a experimental point of view,
to discriminate between a $E(5)$-model and its corresponding IBM
Hamiltonian when only few low-lying states are considered (usually the
four lowest states in the ground state band, plus $0_2^+$ and $2_3^+$
in the $\xi=2$ band).

We have also proved that all the $E(5)-$models correspond to IBM
Hamiltonians very close to the critical area, $|r_1|<0.05$.
Therefore, one can say that the $E(5)-$ models are appropriate to
describe transitional $\gamma-$unstable regions close to the critical
point.

We have found that the results presented in this paper are consistent
with the existence of something similar to a quasidynamical symmetry,
we call this phenomenon quasi-critical point symmetry.

Finally, it should be noted that the use of a more general $U(5)-O(6)$
Hamiltonian, {\it e.g.}~using $\kappa_4$ as free parameter, do not
change the main conclusions of this work.

\section{Acknowledgements}
We are grateful to D.J.~Rowe for a careful reading of the manuscript
and for his valuable comments.  This work has been partially supported
by the Spanish Ministerio de Educaci\'on y Ciencia and by the European
regional development fund (FEDER) under projects number FIS2005-01105,
FPA2006-13807-C02-02 and FPA2007-63074, and by the Junta de
Analuc\'{\i}a under projects FQM160, FQM318, P05-FQM437 and
P07-FQM-02962.


\begin{thebibliography}{25}
\bibitem{Bohr52} A.~ Bohr, Mat.\ Fys.\ Medd.\ Dan.\ Vidensk.\ Selsk.\
  {\bf 26} (14) (1952).
\bibitem{Bohr53}A.~ Bohr, B.R.~ Mottelson, Mat.\ Fys.\ Medd.\ Dan.\
  Vidensk.\ Selsk.\ 27 (16) (1953).
\bibitem{Bohr69} A.~ Bohr, B.R.~ Mottelson, {\it Nuclear Structure},
  vol. II, (Benjamin, Elmsford, NY, 1969).
\bibitem{Arim76} A.~ Arima, F.~ Iachello, Ann.\ Phys.\ {\bf 99}, 253
  (1976).
\bibitem{Arim78} A.~ Arima, F.~ Iachello, Ann.\ Phys.\ {\bf 111}, 201
  (1978);
\bibitem{Scho78} O.~ Scholten, F.~ Iachello, A.~ Arima, Ann.\ Phys.\
  {\bf 115}, 325 (1978);
\bibitem{Iach87} F.~Iachello and A.~Arima, {\em The interacting boson
    model} (Cambridge University Press, Cambridge, 1987).
\bibitem{Jans74} D.~ Janssen, R.V.~ Jolos, F.~ D\"onau, Nucl.\ Phys.\
  A {\bf 224}, 93 (1974).
\bibitem{Diep80a} A.E.L.~ Dieperink, O.~ Scholten, F.~ Iachello,
  Phys.\ Rev.\ Lett.\ {\bf 44}, 1747 (1980).
\bibitem{Diep80b} A.E.L.~ Dieperink and O.~ Scholten, Nucl.\ Phys.\ A
  {\bf 346}, 125 (1980).
\bibitem{Gino80a} J.N.~ Ginocchio, M.W.~ Kirson, Phys.\ Rev.\ Lett.\
  {\bf 44}, 1744 (1980).
\bibitem{Gino80b} J.N.~ Ginocchio, M.W.~ Kirson, Nucl.\ Phys.\ A {\bf
    350}, 31 (1980).
\bibitem{Kirs82} M.W.~ Kirson, Ann.\ Phys.\ (N.Y.) {\bf 143}, 448
  (1982).
\bibitem{Gino82} J.N.~ Ginocchio, Nucl.\ Phys.\ A {\bf 376}, 438
  (1982).
\bibitem{Rowe05} D.J.~ Rowe and G.~ Thiamanova, Nucl.\ Phys.\ A {\bf
    760}, 59 (2005).
\bibitem{Isac99} P.~Van Isacker, Phys.\ Rev.\ Lett.\ {\bf 83}, 4269
  (1999).
\bibitem{Hill53} D.L.~Hill and J.A.~Wheeler, Phys.\ Rev.\ {\bf 89},
  1102, (1953).
\bibitem{Iach00}F.~ Iachello, Phys.\ Rev.\ Lett.\ {\bf 85}, 3580,
  (2000).
\bibitem{Iach01}F.~ Iachello, Phys.\ Rev.\ Lett.\ {\bf 87}, 052502,
  (2001).
\bibitem{Iach03}F.~ Iachello, Phys.\ Rev.\ Lett.\ {\bf 91}, 132502,
  (2003).
\bibitem{Cast00}R.F.~ Casten and N.V.~ Zamfir, Phys.\ Rev.\ Lett.\
  {\bf 85}, 3584, (2000).
\bibitem{Cast01}R.F.~ Casten and N.V.~ Zamfir, Phys.\ Rev.\ Lett.\
  {\bf 87}, 052503 (2001).
\bibitem{GarcXX}J.E.~Garc\'{\i}a-Ramos and J.M.~Arias, in preparation.
\bibitem{Fran01} A.~Frank, C.E.~Alonso, and J.M.~Arias, Phys.\ Rev.\ C
  {\bf 65}, 014301 (2001).
\bibitem{Zamf02} N.V.~Zamfir, {\it et al}, Phys.\ Rev.\ C {\bf 65},
  044325 (2002).
\bibitem{Zhan02} Da-li Zhang and Yu-xin Liu, Phys.\ Rev.\ C {\bf 65},
  057301 (2002).
\bibitem{Aria01} J.M.~Arias, Phys.\ Rev.\ C {\bf 63}, 034308 (2001).
\bibitem{Capr02} M.A.~Caprio, Phys.\ Rev.\ C {\bf 65}, 031304 (2002).
\bibitem{Aria03} J.M.~Arias, C.E. Alonso, A. Vitturi,
  J.E.~Garc\'{\i}a-Ramos, J.~Dukelsky, and A. Frank, Phys.\ Rev.\ C
  {\bf 68}, 041302(R) (2003).
\bibitem{Garc05} J.E.~Garc\'{\i}a-Ramos, J.~Dukelsky, and J.M.~Arias,
  Phys.\ Rev.\ C {\bf 72}, 037301 (2005).
\bibitem{Leva04} G.~ L\'evai and J.M.~ Arias, Phys.\ Rev.\ C {\bf 69},
  014304 (2004).
\bibitem{Ushv94} A.G.~Ushveridze, {\it Quasi-exactly solvable models
    in quantum mechanics} (IOP Publishing, Bristol, 1994).
\bibitem{Bona04} D.~Bonatsos, D.~Lenis, N.~Minkov, P.P.~Raychev, and
  P.A.~Terziev. Phys.\ Rev.\ C {\bf 69}, 044316 (2004).
\bibitem{Fran94} A.~Frank and P.~Van Isacker, {\em Algebraic Methods
    in Molecular and Nuclear Structure Physics} (John Wiley \& Sons,
  NY, 1994).
\bibitem{minuit} F.~James, Minuit: Function Minimization and Error
  Analysis Reference Manual, Version 94.1, CERN, 1994.
\bibitem{Duke84} J.~Dukelsky, G.G.~ Dussel, R.P.J.~ Perazzo, S.L.~
  Reich, and H.M. Sofia, Nucl.\ Phys.\ A {\bf425},93 (1984).
\bibitem{Isac81} P.~Van~Isacker and J.Q.~Chen, Phys.\ Rev.\ C {\bf24},
  684 (1981).
\bibitem{Stan71} H.E.~Standley, ``Introduction to phase transitions
  and critical phenomena'', Oxford University Press, Oxford (1971).
\bibitem{Warn82} D.D.~Warner and R.F.~Casten, Phys.\ Rev.\ Lett.\ {\bf
    48}, 1385 (1982).
\bibitem{Lope96} E.~L\'opez-Moreno and O.~Casta\~nos, \newblock
  {Phys.\ Rev.\ C {\bf 54}, 2374}, (1996).
\bibitem{Gilm81} R.~Gilmore. {\it Catastrophe theory for scientists
    and engineers} (Wiley, New York, 1981).
\bibitem{Aria07} J.M.~Arias, J.~Dukelsky, J.E.~Garc\'{\i}a-Ramos, and
  J.~Vidal, Phys.\ Rev.\ C {\bf 75}, 014301 (2007).
\bibitem{Rowe04a} D.J.~ Rowe, Phys.\ Rev.\ Lett.\ {\bf 93}, 122502
  (2004).
\bibitem{Rowe04b} D.J.~ Rowe, P.S.~ Turner, and G.~ Rosensteel, Phys.\
  Rev.\ Lett.\ {\bf 93}, 232502 (2004).
\bibitem{Rowe04c} D.J.~ Rowe, Nucl.\ Phys.\ A {\bf 745}, 47 (2004).
\bibitem{Rowe05b} P.S.~ Turner and D.J.~ Rowe, Nucl.\ Phys.\ A {\bf
    756}, 333 (2004).
\bibitem{Rowe07} D.J.~ Rowe, private comm.
\end{thebibliography}
\end{document}